# Damping of plasma-electron oscillations and waves in low-collision electron-ion plasmas

V. N. Soshnikov [1]

All-Russian Institute of Scientific and Technical Information
of the Russian Academy of Sciences
(VINITI, Usievitcha 20, 125315 Moscow, Russia)

## Abstract

Previously developed method for finding asymptotic solutions of Vlasov equations using two-dimensional (in coordinate $x$ and time $t$) Laplace transform is applied to low-collision electron-ion plasmas. Taking into account Coulomb collisions in the limit $m_e \ll m_i$, $|v_i|_{mean} \ll |v_e|_{mean}$, and $T_e m_e \ll T_i m_i$ results in the expression for longitudinal high-frequency plasma-electron oscillation wave decrement essentially depending on oscillation frequency $\omega$. This expression is quite different from the used one for low-frequency plasma sound [1, 2, 3, 11, 12], which can be derived using expansion in asymptotically divergent series in $\delta/\omega_0$, $kv_i/\omega_0$, where $\delta$ is imaginary part of the frequency $\omega = \omega_0 + i\delta$ [13], and does not reduce to a simple expression with some effective collision frequency $\delta \sim \nu_{ei}$.

PACS numbers: 52.25 Dg; 52.35 Fp.

Key words: *plasma oscillations; plasma waves; Landau damping; electron waves; Coulomb collisions; collision damping; plasma waves dispersion equation.*

## Introduction

In spite of the fact that Vlasov equations for collisionless plasma appeared more than half a century ago [4], one cannot be satisfied with the problem of solving these equations (and also equations with collision terms). This problem is related with an evident impossibility to obtain asymptotic solutions in the plane-wave form $\exp(i\omega t + ikx)$ with complex frequency $\omega = \omega_0 + i\delta$, independently on the value of real decrement $\delta$. Indeed, the characteristic equation for plasma oscillations (more precisely, its imaginary part when one finds $\delta$) has the following form:

$$\frac{16\pi e^2}{m_e} \int_{-\infty}^{\infty} \partial v_y \int_{-\infty}^{\infty} \partial v_z \int_{0}^{\infty} \partial v_x \frac{\partial f_0}{\partial v_x} \frac{v_x dv_x}{\left(\omega_0^2 - k^2 v_x^2 - \delta^2\right)^2 + 4\delta^2 \omega_0^2} = 0. \quad (1)$$

It is evident, that in the case of background Maxwellian plasma with distribution $\partial f_0/\partial v_x < 0$ this equation has no solutions for any $\delta$ at $\omega_0 \neq 0$, including $\delta \to \pm 0$ (see [5] ).

It is well known that formal substitution of plane-wave solution into Vlasov equations results in appearance of pole divergences in integrals. Correspondingly, integration contour can be arbitrarily selected resulting either in principal value prescription [4] or in analytical continuation to upper or lower half-plane (Landau's continuation produces Landau damping with $\delta > 0$, $\delta \ll \omega_0$, see [6]).

---
[1] Krasnodarskaya str., 51-2-168, Moscow 109559, Russia.

One can relate both selections to some kind of arbitrary solutions which indeed are not the true asymptotic solutions of Vlasov equations. Besides that, in contrary to doubts presented in [5, 7, 8], Landau's analytical continuation itself is based on the classic Cauchy-Riemann theory and is mathematically rigorous. It is interesting to note that one can derive Landau damping solution by expanding formally the integrand in asymptotically divergent series in powers of small parameter $\delta/\omega_0$ and taking into account for only the first non-divergent terms of this expansion [13].

But there are no reasons to believe that asymptotic solutions of Vlasov system of partial differential equations do not exist in principle. A natural approach has been suggested in [7, 8] to search a solution in the form of *a sum of exponents* with amplitudes defined by initial and boundary conditions [2]. It should be emphasized that, due to the connection through initial and boundary conditions, the substitution of this aggregate into the linear Vlasov differential equations does not result in decoupling to the sum of independent equations for *separate* exponents.

Solutions of this type have been obtained in [7, 8] using two-dimensional (in time $t$ and coordinate $x$) Laplace transform of self-consistent plasma electric field $E(x,t) \to E_{p_1 p_2}$. This approach implies searching *a pair of poles* of $E_{p_1 p_2}$, where $p_1 \equiv i\omega$, $p_2 \equiv ik$ are Laplace transform parameters. Amplitudes at exponents are then determined by residues in poles and, for given initial and boundary conditions (see [9, 10]), for example, by additional requirement of solution finiteness at $x,t \to \infty$.

In what follows we present an attempt to extent this method [5, 9, 10] to the case of electron-ion low-collision plasma with heavy and slow ions ( $m_e \ll m_i$, $|\bar{v}_i| \ll |\bar{v}_e|$, and $\omega > \omega_L$ ). In the case of Coulomb collisions we can find solutions using iteration technique: ion and electron velocity distribution functions of perturbed collisionless plasma described by Vlasov equations are used to evaluate Coulomb collision integrals.

## The null iteration

The plasma is assumed to be Maxwellian, homogeneous and infinite in space and time with the boundary condition given in the plane $x = 0$ (the plane geometry). The Vlasov equations for the simplest case of one-dimensional plasma are

$$\frac{\partial f(\vec{v})}{\partial t} + v_x \frac{\partial f(\vec{v})}{\partial x} - \frac{eE}{m_e} \frac{\partial f(\vec{v})}{\partial v_x} = 0; \quad \frac{\partial E}{\partial x} - 4\pi e \int_{-\infty}^{\infty} \Delta n_e d\vec{v} = 0, \tag{2}$$

where $f(\vec{v})$ is the distribution function of charged particles (electrons/ions) normalized to unity

$$f(\vec{v}) = f_0(v) + f_1(\vec{v}, x, t); \quad |f_1| \ll f_0; \tag{3}$$

_______________________________

[2] As an example of such a solution one can imagine standing wave being a sum of two entangled traveling waves.

$f_0(v)$ is Maxwellian distribution, and $\Delta n_e$ is perturbed density of charged particles $(n_e f_1)$. Laplace transforms are introduced according to

$$E(x,t) = \frac{1}{(2\pi i)^2} \int_{\sigma_1-i\infty}^{\sigma_1+i\infty} \int_{\sigma_2-i\infty}^{\sigma_2+i\infty} E_{p_1 p_2} e^{p_1 t} e^{p_2 x} dp_1 dp_2, \tag{4}$$

$$f_1(\vec{v},x,t) = \frac{1}{(2i\pi)^2} \int_{\sigma_1-i\infty}^{\sigma_1+i\infty} \int_{\sigma_2-i\infty}^{\sigma_2+i\infty} f^{(1)}_{p_1 p_2} e^{p_1 t} e^{p_2 x} dp_1 p_2, \tag{5}$$

$$\frac{\partial E(x,t)}{\partial x} = \frac{1}{(2\pi i)^2} \int_{\sigma_1-i\infty}^{\sigma_1+i\infty} \int_{\sigma_2-i\infty}^{\sigma_2+i\infty} p_2 E_{p_1 p_2} e^{p_1 t} e^{p_2 x} dp_1 p_2 - \frac{1}{(2\pi i)^2} \int_{\sigma_1-i\infty}^{\sigma_1+i\infty} \int_{\sigma_2-\infty}^{\sigma_2+i\infty} E_{p_1} e^{p_1 t} e^{p_2 x}, \tag{6}$$

Neglecting for simplicity initial and boundary conditions for $f_1(\vec{v},x,t)$ (they will not affect characteristic frequencies and wave numbers) we obtain

$$p_2 E_{p_1 p_2} = 4\pi e n_e \int_{-\infty}^{\infty} f^1_{p_1 p_2} d\vec{v} + E_{p_1}. \tag{7}$$

After substitution

$$f^{(1)}_{p_1 p_2} = \frac{v_x e}{k_B T_e} \left(\frac{m_e}{2\pi k_B T_e}\right)^{3/2} \exp^{-\frac{m_e v^2}{2 k_B T_e}} \frac{E_{p_1 p_2}}{p_1 + v_x p_2}. \tag{8}$$

one obtains

$$E_{p_1 p_2} = \frac{E_{p_1}}{p_2} + \frac{E_{p_1 p_2}}{p_2} \left(\frac{m_e \omega_L^2}{k_B T_e}\right) \left(\frac{m_e}{2\pi k_B T_e}\right) \int e^{-\frac{m_e v^2}{2 k_B T_e}} \frac{v_x d\vec{v}}{p_1 + v_x p_2} \tag{9}$$

where $\omega_L$ is Langmuir frequency.
Using transformation

$$\int_{-\infty}^{\infty} e^{-\frac{m_e v_x^2}{2 k_B T_e}} \frac{v_x dv_x}{p_1 + v_x p_2} \equiv -2 p_2 \int_0^{\infty} e^{-\frac{m_e v_x^2}{2 k_B T_e}} \frac{v_x^2 dv_x}{p_1^2 - v_x^2 p_2^2} \simeq -\sqrt{\frac{2\pi k_B T_e}{m_e}} \frac{p_2 \overline{v_x^2}}{p_1 - \overline{v_x^2} p_2^2}, \tag{10}$$

where $\overline{v_x^2}$ can be approximated by the mean square velocity defined by Maxwell exponent

$$\overline{v_x^2} \simeq \frac{k_B T_e}{m_e}, \tag{11}$$

one obtains

$$E_{p_1 p_2} \simeq \frac{E_{p_1}}{p_2 \left(1 + \omega_L^2 \big/ \left(p_1^2 - p_2^2 \bar{v}_x^2\right)\right)} \;. \tag{12}$$

Assuming arbitrarily some boundary conditions

$$E(0,t) = E_0 e^{i\omega t} \tag{13}$$

that means

$$E_{p_1} = \frac{E_0}{p_1 - i\omega} \;, \tag{14}$$

one has the pole position in the complex $p_1$ plane

$$p_1 = i\omega \tag{15}$$

and corresponding pole in the complex $p_2$ plane

$$p_2 = \pm i \sqrt{\frac{\omega_L^2 - \omega^2}{\bar{v}_x^2}} \;. \tag{16}$$

At small $n_e$ and $\omega_L^2 < \omega^2$

$$p_2 = ik \simeq \pm i \sqrt{\frac{\omega^2 - \omega_L^2}{\bar{v}_x^2}} \;; \qquad V_{phase} \equiv \frac{\omega}{k} \simeq \frac{\sqrt{\bar{v}_x^2}}{1 - \omega_L^2 / (2\omega^2)} \;, \tag{17}$$

where the later expression is approximately valid at $\omega_L \ll \omega$.
For simplicity we do not discuss here solutions corresponding to some other poles $E_{p_1 p_2}$ in the complex planes of $p_1$ and $p_2$.

## Coulomb low-collision plasma

Using the standard expression for Coulomb collision integrals [11] in the r.h.s. of Vlasov equation (2), substituting therein null-iteration electron and ion distribution functions (8) and integrating by parts one obtains quite cumbersome expressions. These expressions simplify in great extent after neglecting small terms of the order $O(m_e/m_i)$, $O(\bar{v}_i/\bar{v}_e)$, and $O\left[(T_e m_e)/(T_i m_i)\right]$ (low mobile ions) [3]. Then one obtains

---

[3] That means $\dfrac{m_e}{m_i} \ll \dfrac{T_i}{T_e} \ll \dfrac{m_i}{m_e}$

$$E_{p_1 p_2} = \frac{E_{p_1}}{p_2} + \frac{4\pi e^2 n_e}{k_B T_e}\left(\frac{m_e}{2\pi k_B T_e}\right)^{3/2} \frac{E_{p_1 p_2}}{p_2} \int e^{-\frac{m_e v^2}{2 k_B T_e}} \frac{v_x d\vec{v}}{p_1 + v_x p_2}$$

$$+ 8\pi^2 e^6 L \frac{n_e n_i}{m_e^2}\left(\frac{m_e}{2\pi k_B T_e}\right)^{3/2}\left(\frac{m_i}{2\pi k_B T_i}\right)^{3/2} E_{p_1 p_2} \int d\vec{V} e^{-\frac{m_i V^2}{2 k_B T_i}} \times$$

$$\times \int \frac{d\vec{v}}{(p_1 + v_x p_2)^2} \frac{u^2 \delta_{xl} - u_x u_l}{u^3} \frac{\partial}{\partial v_l}\left[\frac{v_x \exp\left(-\frac{m_e v^2}{2 k_B T_e}\right)}{k_B T_e (p_1 + v_x p_2)}\right], \quad (18)$$

where $L$ is the Coulomb logarithm; $u^2 \simeq v^2$, and $u_l \simeq v_l$, where $\vec{u} \equiv \vec{v} - \vec{V}$ and $\vec{V}$ is the ion velocity.

Changing integration region $(-\infty, +\infty)$ in $\int dv_x$ to $(0, +\infty)$, carrying out differentiation $\partial/\partial v_l$ in the second integrand of the r.h.s. of Eq. (18), and approximating $v$, $v^2$, and $v_x^2$ in integrands by their mean values

$$v \to \sqrt{3\overline{v_x^2}}, \qquad v^2 \to 3\overline{v_x^2}, \qquad v_x^2 \to \overline{v_x^2}, \qquad (19)$$

in close analogy to the case of collisionless plasma (see preceding section, Eq. (10)), we arrive at

$$E_{p_1 p_2} \cong \frac{E_0}{p_2 (p_1 - i\omega)} \times \qquad (20)$$

$$\times \frac{\left(p_1^2 - \overline{v_x^2} p_2^2\right)^4}{\left(p_1^2 - \overline{v_x^2} p_2^2\right)^4 + \omega_L^2 \left(p_1^2 - \overline{v_x^2} p_2^2\right)^3 - \frac{16\pi^2 e^6 L p_1 n_e n_i}{3\sqrt{3\overline{v_x^2}} k_B T_e m_e^2}\left[p_1^4 + \left(\overline{v_x^2} p_2^2\right)^2 + 6 p_1^2 \overline{v_x^2} p_2^2\right]},$$

where we used the same boundary condition (13) for $E(0,t)$ as in the preceding section.

We see that at small $\delta$, $|\delta| \ll \sqrt{(\omega^2 - \omega_L^2)/\overline{v_x^2}}$, this function has poles defined by

$$p_1 = i\omega, \qquad (21)$$

$$p_2 = \pm \sqrt{\frac{\omega^2 - \omega_L^2}{\overline{v_x^2}}} + \delta, \qquad (22)$$

with the following expression for coordinate damping decrement:

$$\delta \simeq \pm \frac{8\pi^2 e^6 L n_e n_i}{3\sqrt{3} m_e (k_B T_e)^2} \left[ \frac{8\omega^2 (\omega^2 - \omega_L^2) + \omega_L^4}{\omega_L^6 \sqrt{1 - \omega_L^2/\omega^2}} \right]. \tag{23}$$

The solution with damping in *x* corresponds to a wave traveling in forward direction $(k, \delta < 0)$. Exponentially growing solution (a wave, traveling in backward direction) can be cancelled with a similarly growing term in case of specially selected boundary function $f_1(\vec{v}, 0, t)$ ("self-consistency condition", see [9, 10]).

It is worth to note that according to the notion about traveling waves of the type

$$\exp(i\omega t + ikx + \delta x) = \exp\left(i\omega t + ikx + \delta \frac{x}{t} t\right), \quad \frac{x}{t} = V_{phase} \tag{24}$$

transition from the coordinate decrement to the time decrement is obvious: $\delta_t = V_{phase} \delta_x$. The same expression for $\delta_t$ can be obtained also directly from Eq.(20) assuming that pole $p_1 = i\omega - \delta_t$ with $|\delta_t| \ll \omega$ and at collisionless value of $p_2$ according to Eq.(16).

The above obtained expression for $\delta$ is characterized by the strong frequency dependence and is quite different from the commonly used one for plasma sound with the damping in electron-ion collisions (see [11]) obtained using an expansion in asymptotically divergent series in small parameters $\delta/\omega$, $kv_i/\omega$, including iteration terms for ion-ion collisions [12], and producing as a first (collisionless) approximation Landau damping. It follows from Eq.(20) that decrement is different for different oscillation modes and does not reduce to $\delta_{Landau} + v_{collision}$, where $v_{collision}$ is some static collision frequency independent on $\omega$. And what is more, the expression for $\delta$ include no decrement $\delta_{Landau}$ of Landau damping at all. That one could appear only as a result of addition of a part of the contour integral in complex plane $v_x$, although the integrand at real $v_x$ has now no poles.

It is proposed that one can find in analogous manner parameters of low-frequency plasma-sound modes by adding to Poisson equation in (2) the term $\Delta_i = n_i f_i^1$. As before, collision integral is dominated by the term (18), so the damping will be determined by the electron-ion collision term and can strongly depend on $\omega$.

At $p_1 \equiv i\omega$ taking into account low-frequency plasma sound, plasma wave modes are proposed to be determined as roots of the following equation [9]

$$1 + \frac{\omega_e^2}{p_1^2 - p_2^2 a^2} + \frac{\omega_i^2}{p_1^2 - p_2^2 b^2} = 0 \tag{25}$$

where $\omega_e$ ($\omega_i$) are electron (ion) Langmuir frequencies and $a$ ($b$) are r.m.s. velocities $\sqrt{k_B T_e/m_e}$ ($\sqrt{k_B T_i/m_i}$) (or more generally – some effective velocities),
then for collision plasma one has

$$1 + \frac{\omega_e^2}{p_1^2 - p_2^2 a^2} + \frac{\omega_i^2}{p_1^2 - p_2^2 b^2} - \frac{16\pi^2 e^6 L n_e n_i p_1}{3\sqrt{3} a k_B T_e m_e^2} \times \frac{p_1^4 + p_2^4 a^4 + 6 p_1^2 p_2^2 a^2}{(p_1^2 - p_2^2 a^2)^4} = 0, \tag{26}$$

with $p_2 = p_2^* + \delta$, where $p_2^*$ is a root of Eq.(25), and we assume that $|\delta| \ll |p_2^*|$.

## Conclusions

According to the method of finding plasma oscillation modes with the help of 2-dimensional Laplace transformation [5, 7, 8, 10] we have obtained asymptotical solutions of Vlasov equations for the cases of a collisionless plasma and of a low-collision plasma (with both electron-atom [9] and Coulomb electron-ion collisions, this paper).

Approximations used in this analysis, that is replacements of the type

$$\int f_0(v) w(v_x, v_y, v_z) d\vec{v} \simeq w\left(\sqrt{\overline{v_x^2}}, \sqrt{\overline{v_y^2}}, \sqrt{\overline{v_z^2}}\right) \qquad (27)$$

with specified in this problem function $w$ appear justified, first of all, by the necessity to demonstrate the possibilities of the method in various situations and to find the main, even only qualitative, features of the solutions. By this approach we have obtained in the case $\omega > \omega_L$ a quite new expression (23) for collision electron-waves damping decrement in a low-collision electron-ion plasma with a strong dependence on the frequency $\omega$.

Collision damping with energy dissipation of the Landau damping type is absent as we have shown in [9, 10], and if it will be observed experimentally then it will be related to some possible secondary effects such as the main: non-Maxwellian velocity background, influence of electron collisions with discharge tube walls, wave excitation methods, dispersion, etc.

Any solution of only a single exponent type, $\exp(i\omega t \pm ikx)$, with real or complex $\omega$ and $k$ is not a full solution of the dispersion equation for Vlasov equations which would satisfy all boundary and physical conditions. However one can obtain asymptotical solution as a *sum* of connected exponents with amplitudes defined by initial, boundary and probably some other conditions.

**Acknowledgements.** The author is thankful to Dr. A. P. Bakulev for a help in preparing the first and the following versions of the paper in LaTex style.

*Note to the latest version*

Brief survey with summary of up to day author's works is presented in the last preprint "Logical contradictions of Landau damping", arXiv.org/physics/0610220. Landau's complete (all exponents) solutions on the base of contour integrals do not satisfy physical conditions of absence of backward and kinematical (non-coupled with electrical field) waves in homogeneous plasma half-slab, although they satisfy wave equations. Also $\overline{v_x^2}$ must be replaced with some effective $v_{x\ eff}^2$.